\documentclass[twocolumn,showpacs,preprintnumbers,amsmath,amssymb]{revtex4}

\usepackage{graphicx}
\usepackage{dcolumn}
\usepackage{bm}
\usepackage{subfigure}

\begin{document}

\title{Time-resolved Measurement of Quadrupole Wakefields in Corrugated Structures}
\author{Chao Lu$^{1,2}$}
\author{Feichao Fu$^{1,2}$}
\author{Tao Jiang$^{1,2}$}
\author{Shengguang Liu$^{1,2}$}
\author{Libin Shi$^{1,2}$} 
\author{Rui Wang$^{1,2}$}
\author{Lingrong Zhao$^{1,2}$}
\author{Pengfei Zhu$^{1,2}$}
\author{Zhen Zhang$^{3}$}
\author{Dao Xiang$^{1,2}$}\email{dxiang@sjtu.edu.cn}
\affiliation{$^1$ Key Laboratory for Laser Plasmas (Ministry of Education), Department of Physics and Astronomy, Shanghai Jiao Tong University, Shanghai 200240, China \\ 
$^2$ Collaborative Innovation Center of IFSA (CICIFSA), Shanghai Jiao Tong University, Shanghai 200240, China \\
$^3$ Department of Engineering Physics, Tsinghua University, Beijing, 100084, China}

\date{\today}

\begin{abstract}
Corrugated structures have recently been widely used for manipulating electron beam longitudinal phase space and for producing THz radiation. Here we report on time-resolved measurements of the quadrupole wakefields in planar corrugated structures. It is shown that while the time-dependent quadrupole wakefield produced by a planar corrugated structure causes significant growth in beam transverse emittance, it can be effectively canceled with a second corrugated structure with orthogonal orientation. The strengths of the time-dependent quadrupole wakefields for various corrugated structure gaps are also measured and found to be in good agreement with theories. Our work should forward the applications of corrugated structures in many accelerator based scientific facilities. 

\end{abstract}

\maketitle

\section{INTRODUCTION}

When relativistic electron beams pass through metallic pipes (or plates) with corrugations or dielectric structures, electromagnetic waves (wakefields) that propagate with the beams are excited. Such quasi-single frequency radiation on the one hand is a promising candidate for intense THz source (see, for example \cite{BG,UCLATHz, THzFEL}); on the other hand it may be used to manipulate electron beam longitudinal phase space through the interaction between the electron beam and the electromagnetic waves inside the structure (see, for example \cite{Simpson, Craievich, Bane12}). 

Recently the corrugated structures (CS) and dielectric-based structures have been used to tailor beam longitudinal phase space in three different ways, namely removing linear chirp (correlation between beam's longitudinal position and beam energy), removing quadratic chirp and imprinting energy modulation. For instance, when the electron bunch length is much shorter than the wavelength of the wakefield, the beam sees a decelerating field that increases approximately linearly in longitudinal direction when it passes through a CS. This longitudinal wakefield can be used to ``dechirp'' the beam after bunch compression to reduce beam global energy spread \cite{ Bane12, DC1, DC2, DC3, DC4, DC5}. Alternatively, when the electron bunch length is comparable to the wavelength of the wakefield, the beam sees a longitudinal wake that approximates a sinusoid. This wakefield can be used to compensate for the beam quadratic chirp \cite{ Craievich, Gu, DH, FF2} that otherwise increases free-electron laser (FEL) bandwidth in seeded FELs (see, e.g. \cite{echo15}) and degrades the MeV ultrafast electron microscope (UEM) performance \cite{XD, LR}. Yet another scenario is when electron bunch length is much longer than the wavelength of the wakefield. In this regime the longitudinal wakefield can be used to produce energy modulation in beam longitudinal phase space that may be further converted into density modulation for producing THz radiation \cite{EM1, EM2, EM3}. 
\begin{figure*}[htbp!]
 \centering
  \subfigure{\includegraphics[width=0.95\textwidth]{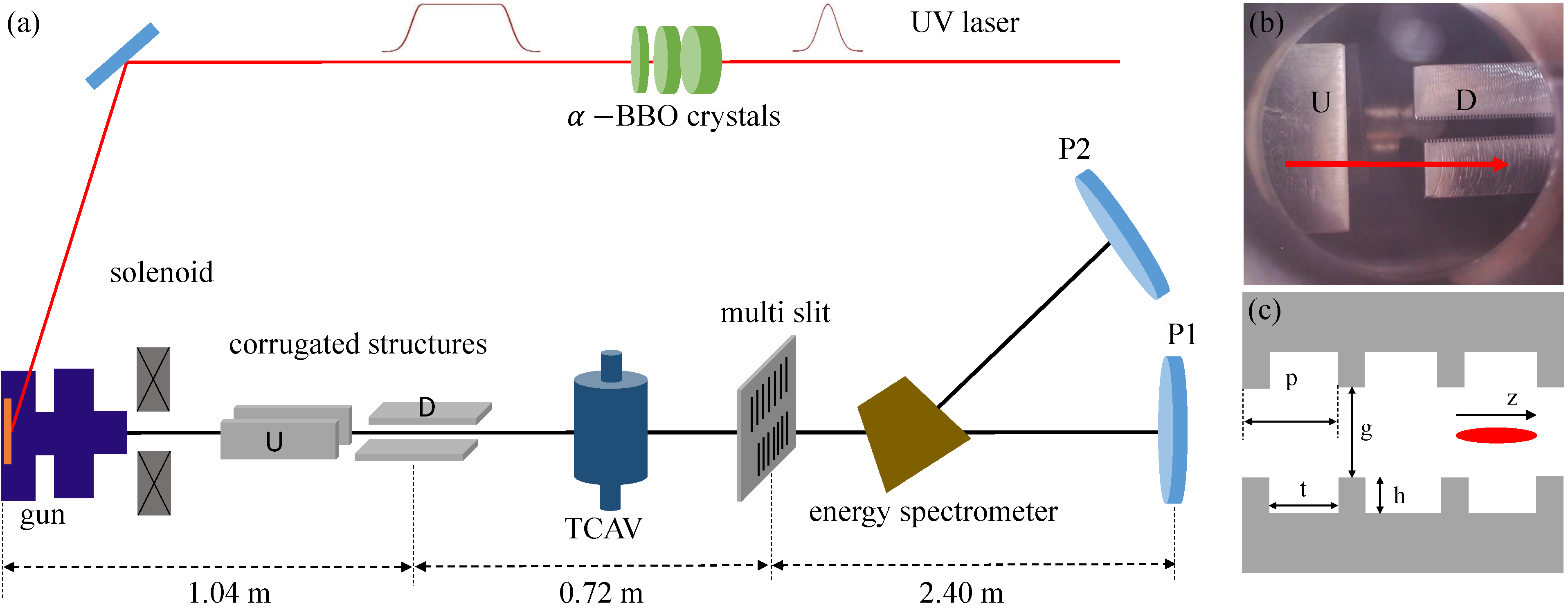}}
\caption{\label{fig1} (color online). (a) Schematic layout of the time-resolved measurement of quadrupole wakefields experiment (distances not shown in scale); (b) side view of the CS: upstream CS has a horizontal gap and downstream CS has a vertical gap (the red arrow indicates the beam direction); (c) geometry of the planar CS parameters (the red ellipse represents a beam propagating along the $z$ axis).}
\end{figure*}

In these applications, the planar CS where the gap can be easily varied to tune the wakefield wavelength and strength is the leading candidate. However, the planar CS also excites quadrupole wakefields even when the beam passes through the structure on-axis. In general the quadrupole wake increases beam transverse emittance by giving beam time-dependent focusing \cite{DC4}. Considering the fact that the transverse wakefields scales as $1/g^4$ ($g$ is the gap of the planar structure) while the longitudinal wake scales as $1/g^2$, this effect if not properly controlled, may lead to serious degradations to beam transverse emittance that might even outweigh the advantage of using this device for manipulating beam longitudinal phase space. Fortunately, analysis shows that the time-dependent quadrupole wakefield produced by a planar CS can be effectively canceled with a second CS with orthogonal orientation \cite{quad}. 

In our previous work \cite{FF2} we demonstrated that the ideas of using CS to removal beam quadratic energy chirp could be applied to the low beam charge (a few pC) and low beam energy (a few MeV) regime, promising for enhancing the performance of MeV UEM. It was also preliminarily shown that the unwanted quadrupole wakes of a pair of CS with orthogonal orientation can be greatly reduced through measurement of the transverse beam size (the beam transverse emittance was not measured). In this paper the beam emittance and transverse phase space in presence of CS are measured with multi-slit method. We show that the time-dependent focusing or defocusing introduced by quadrupole wakefields mainly results in mismatch in beam slice phase space, leading to considerable growth in beam projected emittance. We also provide definitive evidence through measurement of the beam transverse phase space that indeed if the CS is composed of two identical parts with the second half rotated by 90 degrees with respect to the first half, the quadrupole wakes can be canceled. Furthermore, the strengths of the time-dependent quadrupole wakefields for various CS gaps are also measured and found to be in good agreement with theories. We anticipate that our study should forward the applications of CS in many accelerator facilities.

\section{Experiment details and methods to quantify quadrupole wakefields}

The experiment was conducted at the Center for Ultrafast Diffraction and Microscopy at Shanghai Jiao Tong University \cite{FF1}. The schematic layout of the experiment is shown in Fig.~\ref{fig1}(a). By using three alpha barium borate ($\alpha$-BBO) birefringent crystals to stack the UV Gaussian laser into a nearly flat-top pulse, electron beam with approximately uniform distribution (11 ps FWHM) in longitudinal direction is produced (see, e.g. \cite{FF2} and references therein). After exiting the photocathode rf gun, the beam then passes through two planar CS with orthogonal orientations (the upstream CS [U in Fig.~\ref{fig1}(b)] has a horizontal gap and the downstream CS [D in Fig.~\ref{fig1}(b)] has a vertical gap; the upstream and downstream CS are identical parts with the second half rotated by 90 degrees with respect to the first half; more details can be found in \cite{FF2}). Four step motors can be used to change the position of the structures independently, so the gap of the CS as well as the centroid of the gap are both adjustable. The electron beam kinetic energy at the exit of the S-band (2856 MHz) 1.6-cell photocathode rf gun is measured to be about 3.3 MeV with an energy spectrometer. The beam charge is measured to be about 6 pC with a Faraday cup (not shown in Fig.~1) upstream of the CS and the beam longitudinal distribution is measured with an S-band rf transverse cavity (TCAV) downstream of the CS. The gun and TCAV are powered by the same rf station. 

\begin{figure}[b]
 \centering
  \subfigure{\includegraphics[width=0.48\textwidth]{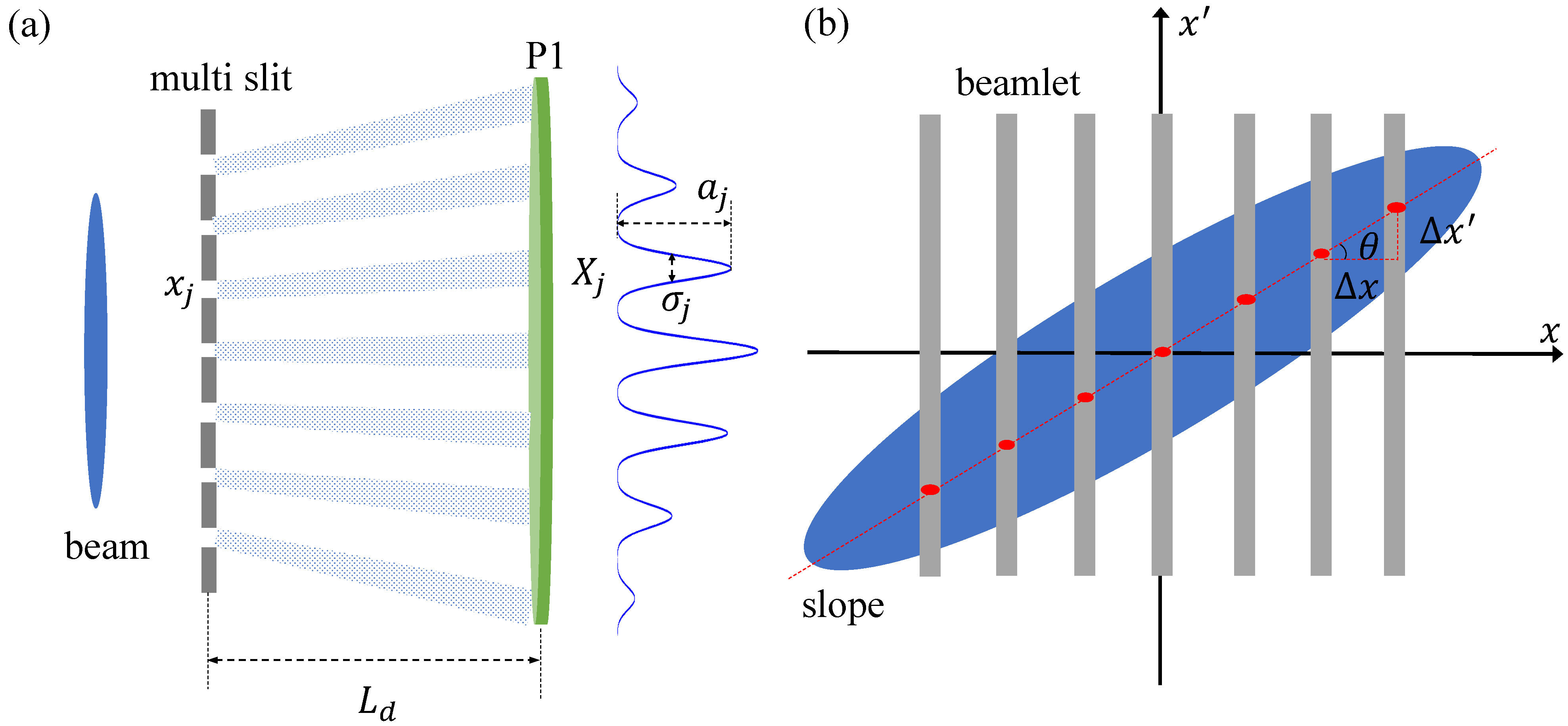}}
\caption{\label{fig2} (color online). (a) Schematic of emittance measurement with multi-slit method; (b) Definition of the ellipse slope.}
\end{figure}

Two stainless steel masks have been used in the experiment to measure the beam transverse emittance through the so-called multi-slit method \cite{ZM}. The recorded image of beamlets is treated as a multi-Gaussian distribution shown in Fig.~\ref{fig2}(a), from which the intensity $a_j$, the position $X_j$ and the rms width $\sigma_j$ of each beamlet ($j =1, 2, ..., p, p$ is the number of slits) can be obtained. The centroid of the beam at the mask is $<x> = \sum_{j=1}^{p} r_jx_j$, where $r_j=\frac{a_j\sigma_j}{\sum_{j=1}^{p} a_j\sigma_j}$ and $x_j$ is the position of the $j$-th slit. The rms size of the sampled beam is calculated as
\begin{equation}\label{xsize}
<x^2> = \sum_{j=1}^{p} r_j (x_j - <x>)^2
\end{equation}

The average divergence of each beamlet is defined as $x_j'=(X_j - x_j)/L_d$ where $L_d$ is the drift length, then the average divergence of the sampled beam is $<x'> = \sum_{j=1}^{p} r_jx_j'$. With the rms divergence of each beamlet being $\sigma_j' = \sigma_j/L_d$, the rms divergence of the sampled beam and the correlation between position and divergence at the mask are found to be

\begin{equation}\label{x'size}
<x'^2> = \sum_{j=1}^{p} r_j [\sigma_j'^2 + (x_j' - <x'>)^2]
\end{equation}

\begin{equation}\label{xx'c}
<xx'> =  \sum_{j=1}^{p} (r_jx_jx_j') - <x><x'>        
\end{equation}

Then the normalized emittance is calculated by 
\begin{equation}\label{emittance}
\varepsilon_x = \gamma \beta \sqrt{<x^2><x'^2> - <xx'>^2} ,
\end{equation}
where $\gamma$ is the relativistic factor and $\beta$ is the beam velocity normalized to the speed of light.

We further define the slope of the phase space ellipse as $S=\tan \theta$ (Fig.~\ref{fig2}(b)). Analysis shows that the slope can be obtained with the multi-slit method and it can be expressed in a very simple formula,
\begin{equation}\label{emittance}
S = \frac{\Delta X-\Delta x}{\Delta x L_d}. 
\end{equation}
where $\Delta X$ is the separation of the beamlets measured at screen P1 and $\Delta x$ is the separation of the slit in the mask.

In this experiment the masks have 7 slits with slit width ($w$) of 85 $\mu$m. The slit separation for the two masks are 350 $\mu$m and 650 $\mu$m, respectively. The drift length $L_d$ from the slits to the phosphor screen (P1) is 2.40 m and the corresponding resolution of the emittance measurement is estimated to be about $\Delta \varepsilon = \gamma\beta w^2/L_d=0.02~\mu$m. In this experiment by streaking the beamlets vertically in screen P1 with the TCAV, time-resolved measurement of the slice emittance can be achieved in a single-shot.

\begin{figure}[b]
 \centering
  \subfigure{\includegraphics[width=0.48\textwidth]{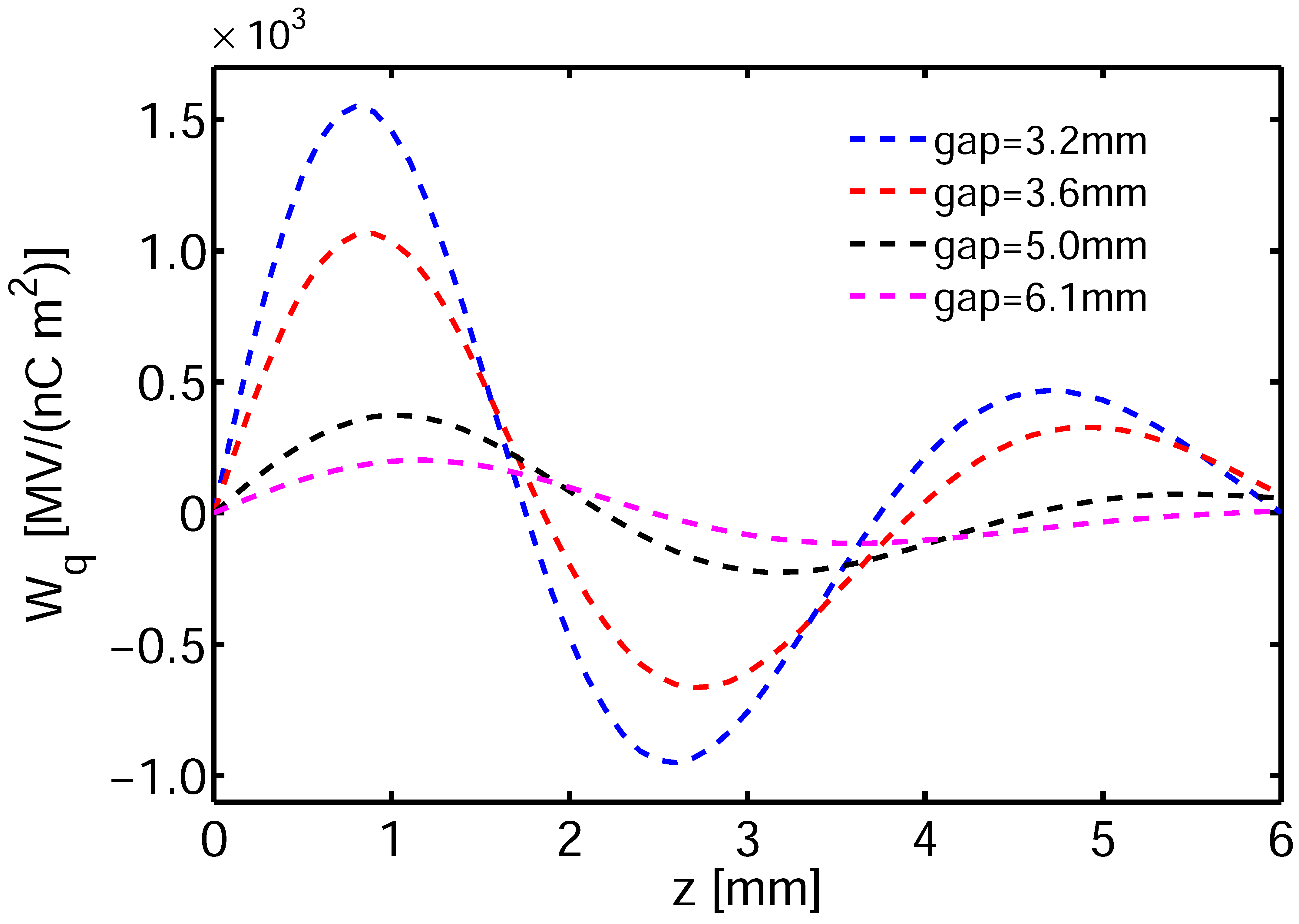}}
\caption{\label{fig3} (color online). The point charge quadrupole wake fields at various CS gaps respectively $g=3.2, 3.6, 5.0, 6.1$ mm.}
\end{figure}

Following the notations in Fig. \ref{fig1}(c), the corrugations are characterized by $h$=0.60 mm, $t$=0.35 mm and $p$=0.6 mm. The lengths of the upstream and downstream CS are both 16 cm. The point charge quadrupole wake fields $W_q(s)$ for various CS gaps calculated using the field matching method (see, e.g. \cite{DC5}) is shown in Fig.~\ref{fig3}. 

The bunch quadrupole wakefield $w_\lambda(z)$ is given by the convolution between the point charge quadrupole wake field and the beam longitudinal distribution,
\begin{equation}\label{eq:eqs}
w_\lambda(z)=\int_{0}^{\infty} W_q(s)\lambda(z-s)\mathrm{d}s
\end{equation} 

The quadrupole wakefield manifests itself as a time-dependent focusing (defocusing) effect and is related to the focal length $f$ as,
\begin{equation}\label{eq:eqs}
1/f(z)=w_\lambda(z)L_c/E
\end{equation} 
where $L_c$ is the length of the CS and $E$ is electron beam energy.

In our experiment the quadrupole wakefields are quantified through measurement of the beam phase space in presence of the CS. When a beam passes through a linear Hamiltonian system, its final beam matrix $\sigma_f$ is connected with its initial beam matrix $\sigma_i$ as $\sigma_f=R\sigma_i R^T$, where $R$ is the symplectic transfer matrix of the system and $R^T$ is transpose of $R$. Under the thin-lens approximation, the Twiss parameters of the beam before ($\alpha_i$, $\beta_i$ and $\gamma_i$) and after ($\alpha_f$, $\beta_f$ and $\gamma_f$) the CS is connected with the CS focusing effect as (see e.g.\cite{MG})

\begin{equation}\label{eq:f}
\begin{bmatrix}\beta_f \\ \alpha_f \\ \gamma_f\end{bmatrix} = \left[
\begin{matrix} 1 & 0 & 0  \\ 1/f  & 1 & 0 \\ 1/f^2 & 2/f & 1 \\ \end{matrix}
\right]
\left[\begin{matrix}\beta_i \\ \alpha_i\\ \gamma_i\end{matrix}\right]
\end{equation}

With the beam Twiss parameters measured with the multi-slit method, one obtains the focal length $f$ with Eq.~(\ref{eq:f}). Inserting $f$ into Eq.~(\ref{eq:eqs}) yields the quadrupole wakefield as 
\begin{equation}\label{eq:ex}
w_\lambda(z)=\frac{E}{L_c}\frac{\alpha_f-\alpha_i}{\beta_i}
\end{equation}

\section{Experimental results}
\subsection{CANCELLATION OF TIME-DEPENDENT QUADRUPOLE WAKE FIELDS}
In this experiment the nearly flat-top electron beam is first streaked vertically in the TCAV; then it passes through the mask with 7 slits, and finally the 7 beamlets are measured at screen P1 after a drift. The TCAV (an rf deflecting structure) gives beam a time-dependent angular kick (i.e.  $y'\propto t$) after passing through at zero-crossing phase. After a drift section the beam angular distribution is converted to spatial distribution, and the vertical axis at the screens (P1 and P2 in Fig. \ref{fig1}(a)) downstream of the TCAV becomes the time axis ($y \propto t$). The mask used here has a slit width of 85 $\mu$m and slit separation of 650 $\mu$m.

\begin{figure}[b]
  \centering
   \subfigure{\includegraphics[width=0.50\textwidth]{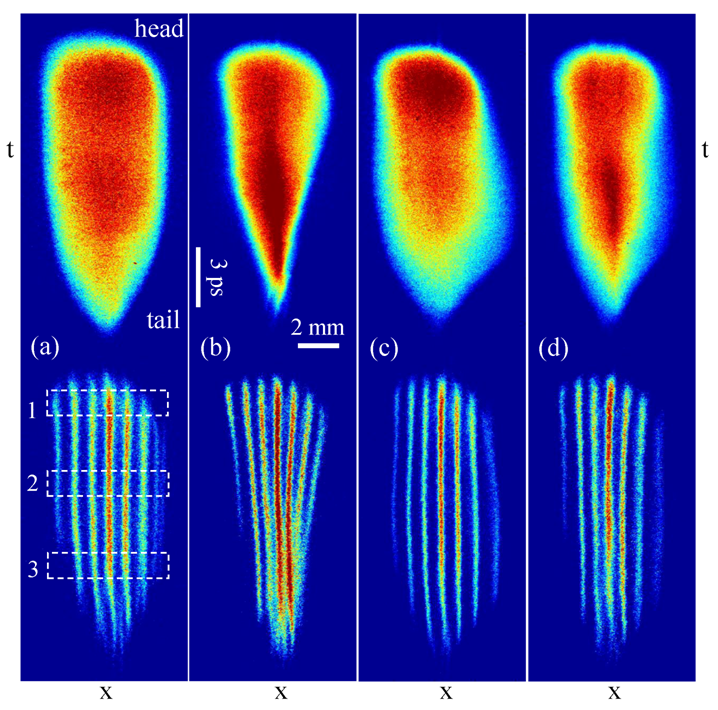}}
\caption{\label{fig4} (color online) Time-resolved measurements of beam (top) and beamlets (bottom) distribution on screen P1. Four situations are shown: (a) with two CS widely open; (b) with downstream CS gap set at 3.2 mm; (c) with upstream CS gap set at 3.2 mm; (d) with both CS both gaps set at 3.2 mm. The dashed squares in (a) indicates the three representative slices used for analysis of emittance and phase space in Fig.~\ref{fig5} and Fig.~\ref{fig6}.}
\end{figure}

With the TCAV on, the streaked beam distributions for various settings of the CS are shown in the top row of Fig.~\ref{fig4}, and the corresponding beamlets distributions with the mask inserted are shown in the bottom row of Fig.~\ref{fig4}. Without going to the details, the quadrupole wake can be qualitatively analyzed. For instance, with the gap of the two CS widely open, the transverse beam size is relatively uniform along the longitudinal direction (the smaller beam size at the bunch tail is attributed to the lower current and thus smaller slice emittance) and the streaked beamlets are parallel to each other (Fig.~\ref{fig4}(a)), indicating a rather well-aligned phase space (e.g. $\Delta X$ is constant from bunch head to bunch tail) for each slice as suggested by Eq.~(\ref{emittance}). With the gap of the downstream CS reduced to 3.2 mm, the beam distribution is shown in Fig.~\ref{fig4}(b), where one can see that the wakefield produced by the head of the beam strongly focuses the tail of the beam such that the bunch tail horizontal size is significantly reduced. Because of this time-dependent focusing effect, the separation of the beamlets is reduced from head to tail, indicating clockwise rotation and mismatch of the slice phase space. It should be pointed out that this quadrupole wakefield should defocus the beam in vertical direction. However, because the beam is also streaked in vertical direction, the defocusing effect is hard to see here. 

Similarly, with the upstream CS reduced to 3.2 mm (the downstream structure is open), the time-dependent quadrupole wakefield defocuses the tail part of the beam, as shown in Fig.~\ref{fig4}(c). As a result of this time-dependent defocusing, the separation of the beamlets is increased from head to tail, corresponding to counterclockwise rotation of the slice phase space. Finally, with the gaps of the two CS both set at 3.2 mm, the quadrupole wakefields effectively cancel each other and the beam distribution (Fig.~\ref{fig4}(d)) is similar to the situation without the CS. 

\begin{figure}[b]
 \centering
   \subfigure{\includegraphics[width=0.50\textwidth]{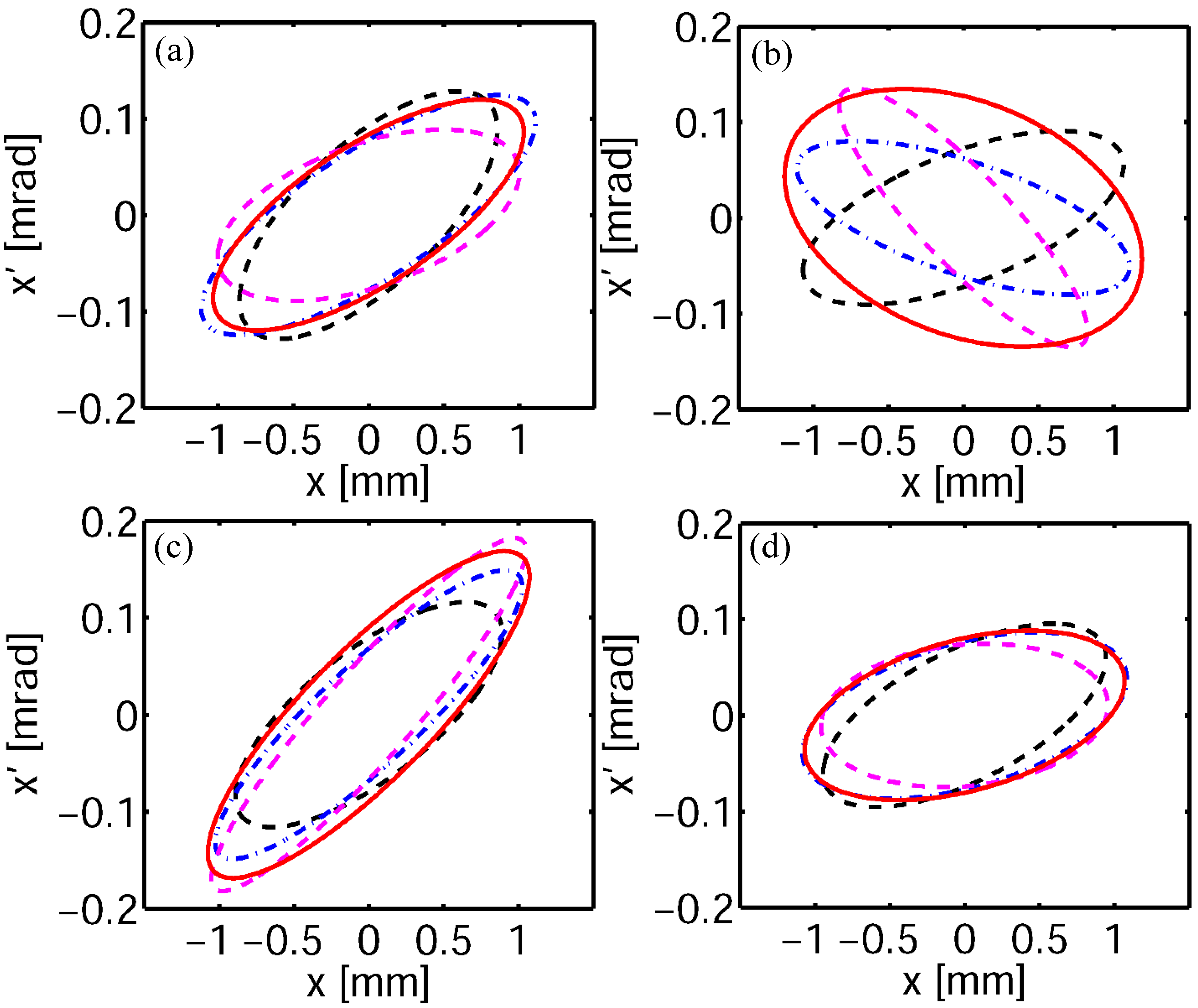}}
\caption{\label{fig5} Slice and projected phase space ellipses in four situations respectively: (a) with two CS widely open; (b) with downstream CS gap set at 3.2 mm; (c) with upstream CS gap set at 3.2 mm; (d) with both CS gaps set at 3.2 mm. The phase spaces for the head slice, central slice and tail slice are shown with dashed black line, dotted dashed blue line and dashed magenta line, respectively. The projected phase space is shown with solid red line.}
\end{figure}

\begin{figure}[b]
   \includegraphics[width=0.50\textwidth]{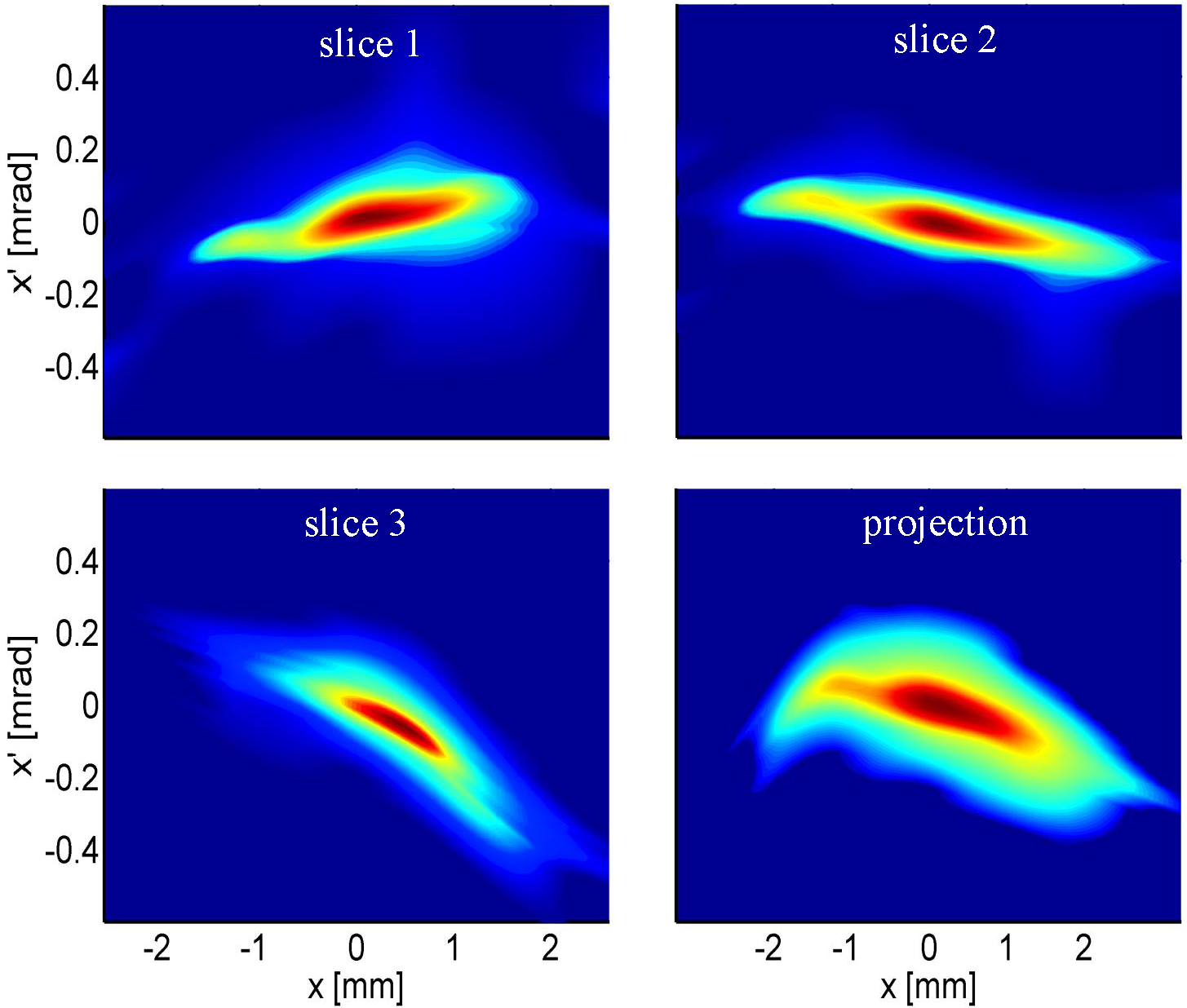}
\caption{\label{fig6} Reconstructed phase space distributions for the three representative slices and the whole beam with the gap of the downstream CS set at 3.2 mm.}
\end{figure}

To give quantitative analysis of the quadrupole wakefield and its effect on beam emittance, the streaked beamlets may be divided into many slices and the phase space for each slice can be obtained with the multi-slit method. Here we give the results for three representative slices (bunch head, bunch center and bunch tail) as indicated by the dashed squares in Fig.~\ref{fig4}(a). The phase spaces for the representative slices under various conditions (Fig.~\ref{fig4}) are shown in Fig.~\ref{fig5}. The slice emittance (throughout this paper normalized emittance values are quoted) is about 0.55 $\mu$m and the projected emittance is about 0.6 $\mu$m when the two CS are widely open (Fig.~\ref{fig5}(a)). Because the phase space ellipse for each slice has approximately the same slope, the projected beam emittance is close to the slice emittance. It should be noted that while due to the weaker space charge force at the bunch tail (current is low in this region) the phase space ellipse for the slice at the bunch tail has a slightly smaller slope (dashed magenta line in Fig.~\ref{fig5}(a)), it does not significantly affect the projected emittance because the number of particles in this region is also small. With the gap of the downstream CS reduced to 3.2 mm, it can be clearly seen that slice phase space rotates clockwise from head to tail (Fig.~\ref{fig5}(b)). As a result of this mismatch, the projected beam emittance increases to about 1.0 $\mu$m (note, the slice emittance is reduced to about 0.50 $\mu$m because a small fraction of the particles (about 15\%) with large offset are collimated by the CS). For this case we also give the reconstructed phase space for the three representative slices and for the whole beam as shown in Fig.~\ref{fig6}. 

Similarly, when the gap of the upstream CS is reduced to 3.2 mm, the time-dependent defocusing wake rotates the slice phase space counter clockwise (Fig.~\ref{fig5}(c)). It is worth mentioning that in Fig.~\ref{fig5}(c) the projected emittance is measured to be about 0.70 $\mu$m, smaller than that in Fig.~\ref{fig5}(b). This is likely due to the fact that the horizontal beta function is smaller in the upstream CS (beam is diverging) such that the emittance growth is less sensitive to the quadrupole wake \cite{quad}. Finally, with the gaps of the two CS both set at 3.2 mm, the slice phase ellipses becomes approximately aligned again as shown in Fig.~\ref{fig5}(d). Because of the effective cancellation of quadrupole wakefields, the projected beam emittance is reduced to about 0.60 $\mu$m. Note, because of the different beta functions in the upstream and downstream CS, there is still a slight increase of the beam emittance even though the quadrupole wake has essentially the same strength and opposite sign. In practical cases, one should take into account the difference of the beta functions in the structures and adjust the gap of the structures accordingly to fully cancel the emittance growth. 

\begin{figure}[b]
 \centering
   \subfigure{\includegraphics[width=0.46\textwidth]{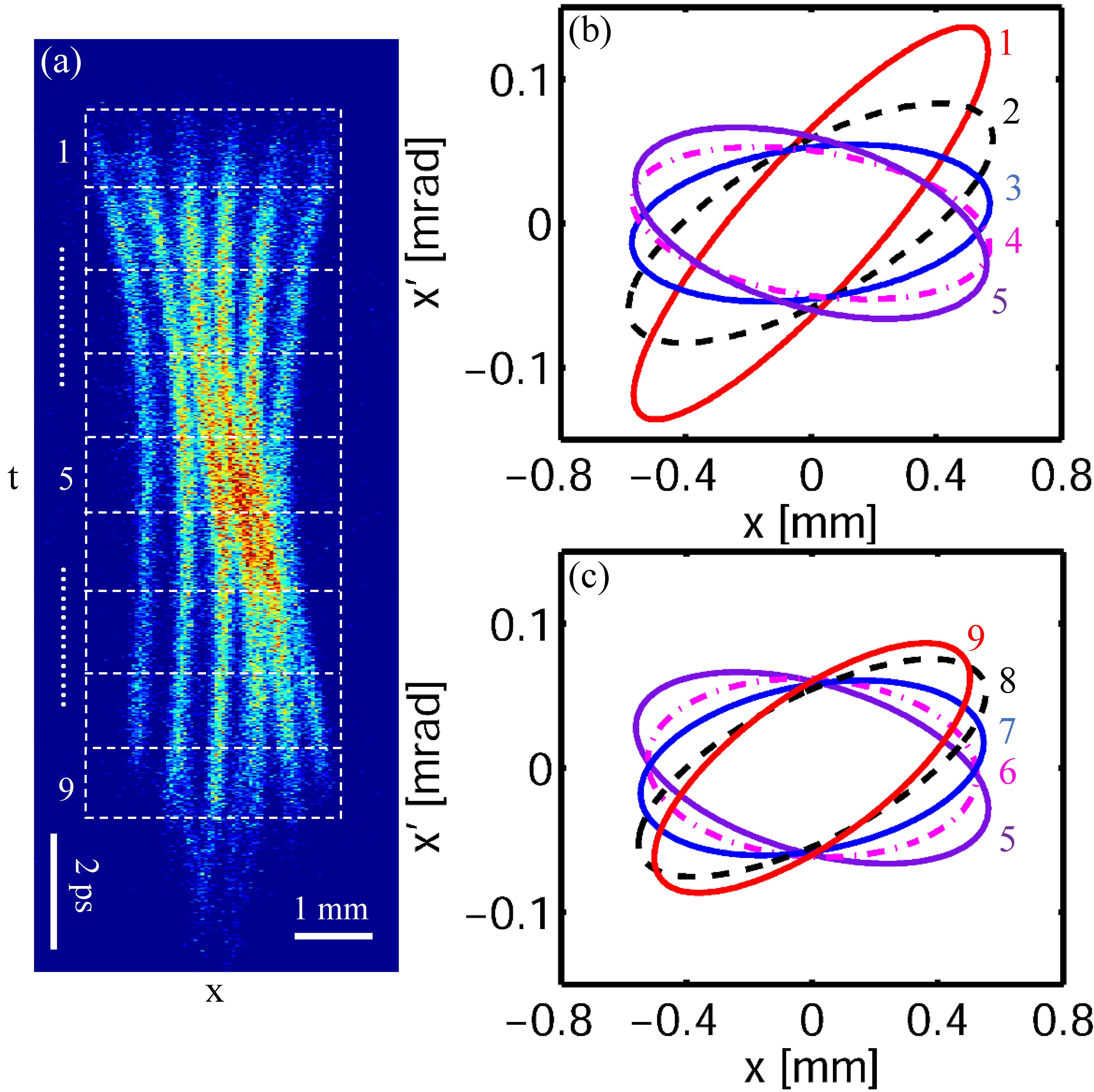}} 
\caption{\label{fig7} (a) Streaked beam distribution with the gap of the downstream CS reduced to 1.4 mm; (b) Phase space ellipses for the front five slices (from bunch head to bunch center); (c) Phase space ellipses for the latter five slices (from bunch center to bunch tail).}
\end{figure}

In a separate experiment, the gap of the downstream CS was further reduced to about 1.4 mm. In this case a large fraction of the electrons are collimated by the structure and only about 2 pC charge went through. The charge is inferred from the intensity of the beam image at screen P1 where the intensity is calibrated with the Faraday cup without the CS. The streaked beamlets are shown in Fig.~\ref{fig7}(a). In this case we used the second mask of which the separation of the slit is 350 $\mu$m. At first glance one might think the quadrupole wake focuses the beam at the head and defocuses it at the tail. However, after second thought one realizes that actually the beam is focused by the wake all the way from head to tail because the horizontal beam size is always smaller than that at the bunch head. Because the focusing is strongest in the beam center (the beamlets almost overlap in the beam center), the beamlets finally developed to a curved shape. 

To clearly show the relative orientation of the slice phase space ellipses, the beam is divided into 9 slices as indicated in Fig.~\ref{fig7}(a). As shown in Fig.~\ref{fig7}(b) the front five slice phase ellipses rotate clockwise because the focusing strength increases from the bunch head to the bunch center. After reaching the peak value at the bunch center, the focusing strength of the wakefield decreases from bunch center to bunch tail, and as a result the last five slice phase ellipses rotate counterclockwise as shown in Fig.~\ref{fig7}(c). The slice emittance is measured to be about 0.20 $\mu$m (primarily due to the reduced charge) and the projected emittance is about 0.40 $\mu$m. 

It is worth mentioning that the emittance grows here by only a factor of 2 even though the gap is reduced to 1.4 mm. This is because the beam charge is also low in this case (wakefield strength is proportional to beam charge). Furthermore,  a mirror symmetry (i.e. the bunch tail is similar to the mirror image of the bunch head with respect to the bunch center) is developed in the final beam distribution because the time-dependent quadrupole wake peaks at the bunch center. This reduces the mismatch of the slice phase space and thus leads to reduced emittance growth (e.g. the phase space ellipses for slice 1 and slice 9 have roughly the same slope; similarly slice 2 and slice 8 are also aligned, etc.).

\subsection{QUADRUPOLE WAKE FIELDS AT VARIOUS LONGITUDINAL POSITIONS}

In addition to studying the effect of quadrupole wakefields on beam emittance growth, here we also present the measurement of the quadruple wakefield strength. By measuring the phase space for various slices one obtains the Twiss parameters of each slice and the quadruple wakefield strengths at various longitudinal positions can be quantified with Eq.~(\ref{eq:ex}). It should be pointed out that with the multi-slit method it is the beam phase space at the mask that is measured. In our analysis the beam phase space is back propagated to the center of the CS to obtain the Twiss parameters at the CS. Specifically, the Twiss parameters at the CS obtained without the CS is taken as those ``before'' the CS ($\alpha_i$, $\beta_i$ and $\gamma_i$ in Eq.~(\ref{eq:f})), and the Twiss parameters obtained with the CS is taken as those ``after'' the CS ($\alpha_f$, $\beta_f$ and $\gamma_f$ in Eq.~(\ref{eq:f})).

\begin{figure}[t]
\includegraphics[width=0.50\textwidth]{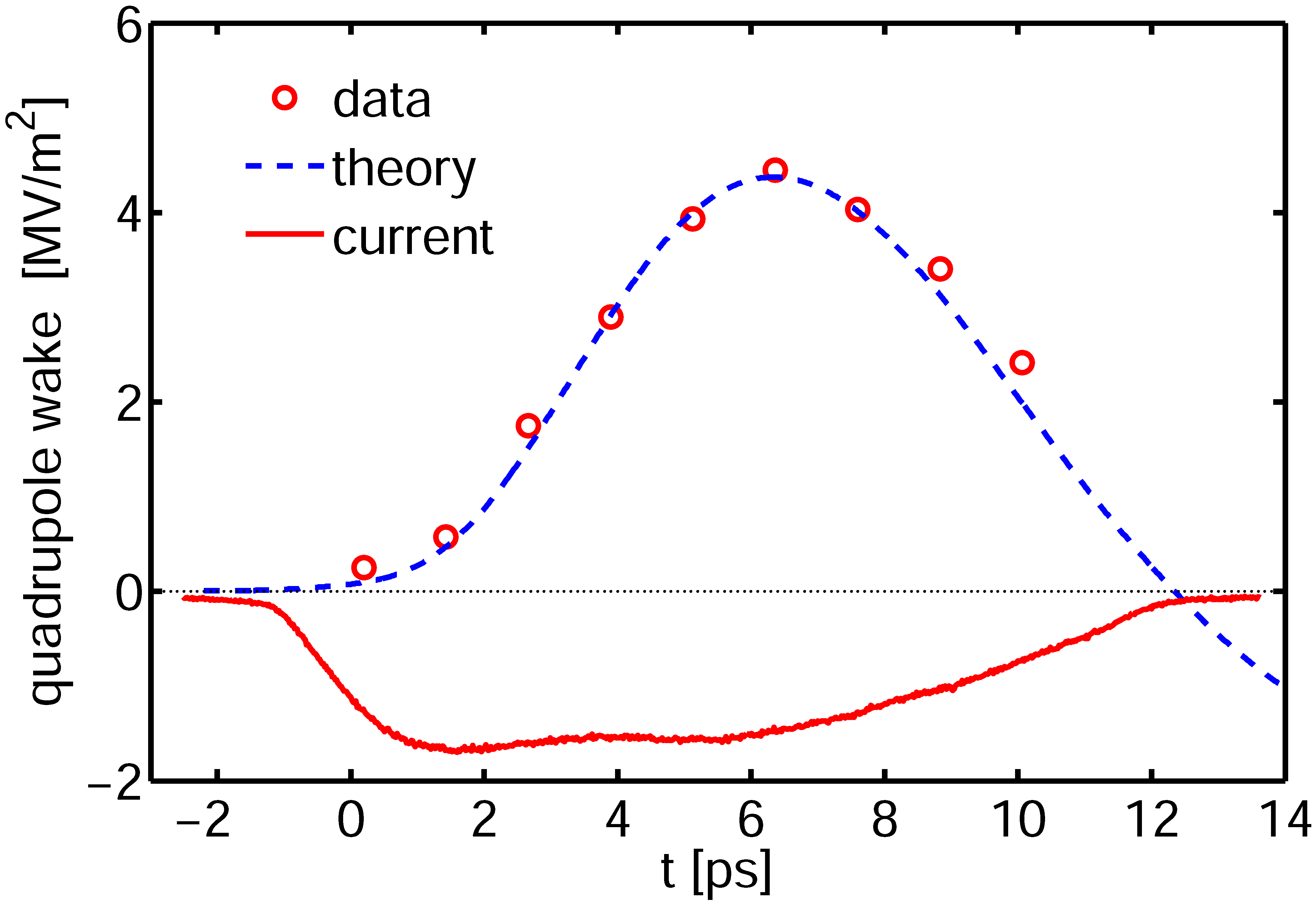}
\caption{\label{fig8} Measured (red circles) and simulated (blue dashed line) time-dependent quadrupole wakefields with the gap of downstream CS set at 3.2 mm. The beam longitudinal distribution is also shown in red solid line.}
\end{figure}

With this method the time-dependent quadrupole wakefield for the beam in Fig.~\ref{fig4}(b) is shown in Fig.~\ref{fig8}. The simulated quadrupole wakefield obtained by convolving the beam longitudinal distribution (red line in Fig.~\ref{fig8}) with the point charge quadrupole wake field is also shown in Fig.~\ref{fig8} with the blue dashed line. To match the experimental results, in the simulation the bunch charge is assumed to be 8 pC which is slightly higher than the measured beam charge. This may be due to the fact that the point wake used for analysis (Fig.~3) only considers the dominate modes of the wakefield, and there might be contributions from other modes \cite{Sasha}. Also there could be considerable uncertainty in measurement of the beam charge with our home-made Faraday cup where secondary electron emission is not minimized. Nevertheless, the overall shape and strength of the quadrupole wakefield is in good agreement with the experimental results.  

\begin{figure}[b]
\includegraphics[width=0.5\textwidth]{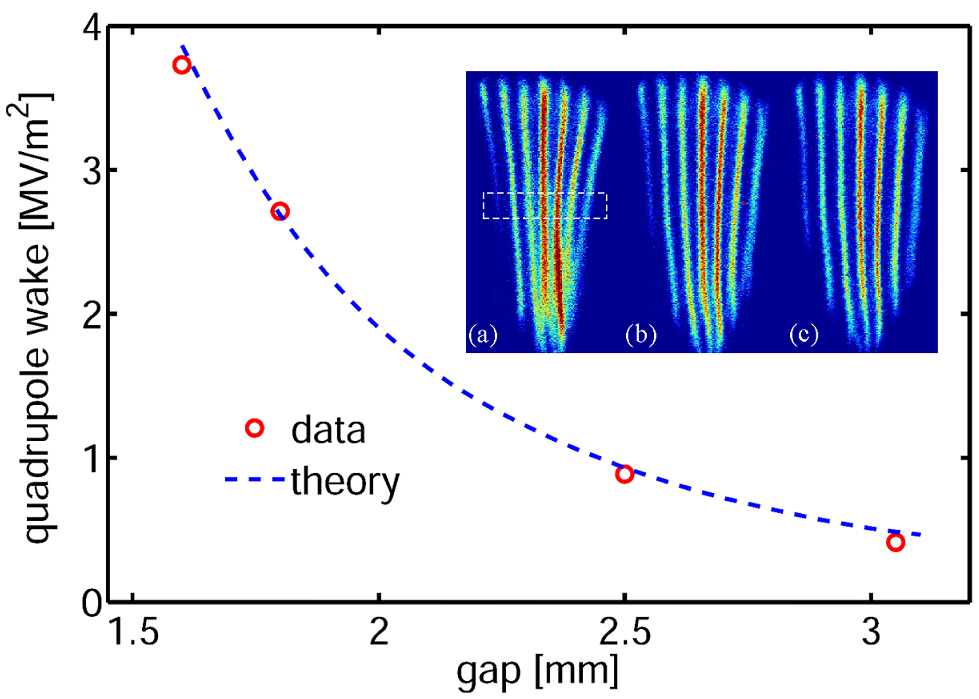}
\caption{\label{fig9} Measured (red circles) and simulated (blue dashed line) quadrupole wakefields of the central slice at various gaps. The streaked beamlets at various CS gaps are shown in the insets [(a) for $g=3.6$ mm, (b) for $g=5.0$ mm and (c) for $g=6.1$ mm].The dashed square indicates the region of the central slice used for analysis .}
\end{figure}

From Fig.~\ref{fig8}, one finds that the maximal quadruple wakefield is about 4.5 MV/m$^2$, corresponding to a focal length of about 5 m much larger than the length of the CS. So the assumption of thin-lens for the CS and the analysis in Eq.~(\ref{eq:f}) is well justified. It should also be pointed out that for the case when CS gap is reduced to 1.4 mm (Fig.~\ref{fig7}), while the Twiss parameters of the beam ``after'' the CS can be obtained, those ``before'' the CS are hard to obtain in experiment because of the considerable beam loss in the CS. Furthermore, it is not clear if the beam gradually loses charge during passage through the CS or most of the beam loss occur at the entrance of the CS. This gives large uncertainty in the charge that contributes to quadrupole wakefield and thus the quadrupole wakefield strength for the 1.4 mm gap case cannot be accurately quantified in this experiment.

\subsection{QUADRUPOLE WAKE FIELDS AT VARIOUS CS GAPS}
To study how the quadrupole wakefield scales with the CS gap, we increased the CS gap to 3.6, 5.0 and 6.1 mm. The measured beamlets are shown in the inset of Fig.~\ref{fig9}. The quadrupole wakefield strength at the central slice for various CS gaps are obtained through measurement of the Twiss parameters of the central slice. The beam loss is about 5\% for $g=3.6$ mm and is negligible for the other two cases, so the beam longitudinal distribution may be assumed identical. The measured quadrupole wake at various CS gap is normalized to the beam charge ($Q=5$ pC as for the $g=3.2$ mm case) and shown with red circles in Fig.~\ref{fig9}. The theoretical wakefield strength is then similarly obtained by a convolution of the beam distribution (assuming $Q=8$ pC) with the point wake, and is found to be in good agreement with the experimental results.

\section{CONCLUSIONS}
In summary, we provided a complete characterization of the quadrupole wakefield in planar CS. It is demonstrated that while the time-dependent quadrupole wakefield produced by a planar CS causes significant growth in beam transverse emittance, it can be effectively canceled with a second CS with orthogonal orientation. Our work should forward the applications of CS in many accelerator based scientific facilities.
  
\section{ACKNOWLEDGMENTS}
This work was supported by the Major State Basic Research Development Program of China (Grants No. 2015CB859700) and by the National Natural Science Foundation of China (Grants No. 11327902). One of the authors (DX) would like to thank the support from the Program for Professor of Special Appointment (Eastern Scholar) at Shanghai Institutions of Higher Learning (No.SHDP201507).

\end{document}